\documentclass[aps,prl,twocolumn,showpacs,amsmath,amssymb,bm,superscriptaddress,reprint,nofootnoteinbib]{revtex4-1}  

\usepackage{graphicx}  
\usepackage{dcolumn}   
\usepackage{hyperref}
\usepackage{bm}        
\usepackage{subfigure}

\newcommand{\Dsm}{\ensuremath{D_s^-}}
\newcommand{\Dsp}{\ensuremath{D_s^+}}
\newcommand{\Ds}{\ensuremath{D_s^\pm}}

\newcommand{\Dsdecay}{\ensuremath{D_s^\pm \rightarrow \phi \pi^{\pm}}}

\newcommand{\DsMdecay}{\ensuremath{D_s^- \rightarrow \phi \pi^{-}}}

\newcommand{\Bsdecay}{\ensuremath{B_s^0 \rightarrow \mu^+ D_s^-}}

\newcommand {\Bs} {\ensuremath{B^0_s}}

\newcommand {\barBs} {\ensuremath{\bar{B}^0_s}}

\begin{document}

\hspace{5.2in} \mbox{FERMILAB-PUB-15-258-E}

\title{
Search for Violation of  $\bm{CPT}$ and Lorentz invariance in
$\bm{B_s^0}$ meson oscillations}
\affiliation{LAFEX, Centro Brasileiro de Pesquisas F\'{i}sicas, Rio de Janeiro, Brazil}
\affiliation{Universidade do Estado do Rio de Janeiro, Rio de Janeiro, Brazil}
\affiliation{Universidade Federal do ABC, Santo Andr\'e, Brazil}
\affiliation{University of Science and Technology of China, Hefei, People's Republic of China}
\affiliation{Universidad de los Andes, Bogot\'a, Colombia}
\affiliation{Charles University, Faculty of Mathematics and Physics, Center for Particle Physics, Prague, Czech Republic}
\affiliation{Czech Technical University in Prague, Prague, Czech Republic}
\affiliation{Institute of Physics, Academy of Sciences of the Czech Republic, Prague, Czech Republic}
\affiliation{Universidad San Francisco de Quito, Quito, Ecuador}
\affiliation{LPC, Universit\'e Blaise Pascal, CNRS/IN2P3, Clermont, France}
\affiliation{LPSC, Universit\'e Joseph Fourier Grenoble 1, CNRS/IN2P3, Institut National Polytechnique de Grenoble, Grenoble, France}
\affiliation{CPPM, Aix-Marseille Universit\'e, CNRS/IN2P3, Marseille, France}
\affiliation{LAL, Universit\'e Paris-Sud, CNRS/IN2P3, Orsay, France}
\affiliation{LPNHE, Universit\'es Paris VI and VII, CNRS/IN2P3, Paris, France}
\affiliation{CEA, Irfu, SPP, Saclay, France}
\affiliation{IPHC, Universit\'e de Strasbourg, CNRS/IN2P3, Strasbourg, France}
\affiliation{IPNL, Universit\'e Lyon 1, CNRS/IN2P3, Villeurbanne, France and Universit\'e de Lyon, Lyon, France}
\affiliation{III. Physikalisches Institut A, RWTH Aachen University, Aachen, Germany}
\affiliation{Physikalisches Institut, Universit\"at Freiburg, Freiburg, Germany}
\affiliation{II. Physikalisches Institut, Georg-August-Universit\"at G\"ottingen, G\"ottingen, Germany}
\affiliation{Institut f\"ur Physik, Universit\"at Mainz, Mainz, Germany}
\affiliation{Ludwig-Maximilians-Universit\"at M\"unchen, M\"unchen, Germany}
\affiliation{Panjab University, Chandigarh, India}
\affiliation{Delhi University, Delhi, India}
\affiliation{Tata Institute of Fundamental Research, Mumbai, India}
\affiliation{University College Dublin, Dublin, Ireland}
\affiliation{Korea Detector Laboratory, Korea University, Seoul, Korea}
\affiliation{CINVESTAV, Mexico City, Mexico}
\affiliation{Nikhef, Science Park, Amsterdam, the Netherlands}
\affiliation{Radboud University Nijmegen, Nijmegen, the Netherlands}
\affiliation{Joint Institute for Nuclear Research, Dubna, Russia}
\affiliation{Institute for Theoretical and Experimental Physics, Moscow, Russia}
\affiliation{Moscow State University, Moscow, Russia}
\affiliation{Institute for High Energy Physics, Protvino, Russia}
\affiliation{Petersburg Nuclear Physics Institute, St. Petersburg, Russia}
\affiliation{Instituci\'{o} Catalana de Recerca i Estudis Avan\c{c}ats (ICREA) and Institut de F\'{i}sica d'Altes Energies (IFAE), Barcelona, Spain}
\affiliation{Uppsala University, Uppsala, Sweden}
\affiliation{Taras Shevchenko National University of Kyiv, Kiev, Ukraine}
\affiliation{Lancaster University, Lancaster LA1 4YB, United Kingdom}
\affiliation{Imperial College London, London SW7 2AZ, United Kingdom}
\affiliation{The University of Manchester, Manchester M13 9PL, United Kingdom}
\affiliation{University of Arizona, Tucson, Arizona 85721, USA}
\affiliation{University of California Riverside, Riverside, California 92521, USA}
\affiliation{Florida State University, Tallahassee, Florida 32306, USA}
\affiliation{Fermi National Accelerator Laboratory, Batavia, Illinois 60510, USA}
\affiliation{University of Illinois at Chicago, Chicago, Illinois 60607, USA}
\affiliation{Northern Illinois University, DeKalb, Illinois 60115, USA}
\affiliation{Northwestern University, Evanston, Illinois 60208, USA}
\affiliation{Indiana University, Bloomington, Indiana 47405, USA}
\affiliation{Purdue University Calumet, Hammond, Indiana 46323, USA}
\affiliation{University of Notre Dame, Notre Dame, Indiana 46556, USA}
\affiliation{Iowa State University, Ames, Iowa 50011, USA}
\affiliation{University of Kansas, Lawrence, Kansas 66045, USA}
\affiliation{Louisiana Tech University, Ruston, Louisiana 71272, USA}
\affiliation{Northeastern University, Boston, Massachusetts 02115, USA}
\affiliation{University of Michigan, Ann Arbor, Michigan 48109, USA}
\affiliation{Michigan State University, East Lansing, Michigan 48824, USA}
\affiliation{University of Mississippi, University, Mississippi 38677, USA}
\affiliation{University of Nebraska, Lincoln, Nebraska 68588, USA}
\affiliation{Rutgers University, Piscataway, New Jersey 08855, USA}
\affiliation{Princeton University, Princeton, New Jersey 08544, USA}
\affiliation{State University of New York, Buffalo, New York 14260, USA}
\affiliation{University of Rochester, Rochester, New York 14627, USA}
\affiliation{State University of New York, Stony Brook, New York 11794, USA}
\affiliation{Brookhaven National Laboratory, Upton, New York 11973, USA}
\affiliation{Langston University, Langston, Oklahoma 73050, USA}
\affiliation{University of Oklahoma, Norman, Oklahoma 73019, USA}
\affiliation{Oklahoma State University, Stillwater, Oklahoma 74078, USA}
\affiliation{Brown University, Providence, Rhode Island 02912, USA}
\affiliation{University of Texas, Arlington, Texas 76019, USA}
\affiliation{Southern Methodist University, Dallas, Texas 75275, USA}
\affiliation{Rice University, Houston, Texas 77005, USA}
\affiliation{University of Virginia, Charlottesville, Virginia 22904, USA}
\affiliation{University of Washington, Seattle, Washington 98195, USA}
\author{V.M.~Abazov} \affiliation{Joint Institute for Nuclear Research, Dubna, Russia}
\author{B.~Abbott} \affiliation{University of Oklahoma, Norman, Oklahoma 73019, USA}
\author{B.S.~Acharya} \affiliation{Tata Institute of Fundamental Research, Mumbai, India}
\author{M.~Adams} \affiliation{University of Illinois at Chicago, Chicago, Illinois 60607, USA}
\author{T.~Adams} \affiliation{Florida State University, Tallahassee, Florida 32306, USA}
\author{J.P.~Agnew} \affiliation{The University of Manchester, Manchester M13 9PL, United Kingdom}
\author{G.D.~Alexeev} \affiliation{Joint Institute for Nuclear Research, Dubna, Russia}
\author{G.~Alkhazov} \affiliation{Petersburg Nuclear Physics Institute, St. Petersburg, Russia}
\author{A.~Alton$^{a}$} \affiliation{University of Michigan, Ann Arbor, Michigan 48109, USA}
\author{A.~Askew} \affiliation{Florida State University, Tallahassee, Florida 32306, USA}
% ...
%
%
\collaboration{The D0 Collaboration\footnote{with visitors from
$^{a}$Augustana College, Sioux Falls, SD, USA,
$^{b}$The University of Liverpool, Liverpool, UK,
$^{c}$DESY, Hamburg, Germany,
$^{d}$CONACyT, Mexico City, Mexico,
$^{e}$SLAC, Menlo Park, CA, USA,
$^{f}$University College London, London, UK,
$^{g}$Centro de Investigacion en Computacion - IPN, Mexico City, Mexico,
$^{h}$Universidade Estadual Paulista, S\~ao Paulo, Brazil,
$^{i}$Karlsruher Institut f\"ur Technologie (KIT) - Steinbuch Centre for Computing (SCC),
D-76128 Karlsruhe, Germany,
$^{j}$Office of Science, U.S. Department of Energy, Washington, D.C. 20585, USA,
$^{k}$American Association for the Advancement of Science, Washington, D.C. 20005, USA,
$^{l}$Kiev Institute for Nuclear Research, Kiev, Ukraine,
$^{m}$University of Maryland, College Park, Maryland 20742, USA
and
$^{n}$European Orgnaization for Nuclear Research (CERN), Geneva, Switzerland
}} \noaffiliation
\vskip 0.25cm

\date{June 12, 2015}

\begin{abstract}
	We present the first search for  CPT-violating effects in the mixing
	of \Bs\ mesons using the full Run II data set with an integrated
	luminosity  of 10.4 fb$^{-1}$ of proton-antiproton collisions
	collected using the D0 detector at the Fermilab Tevatron Collider.
	We measure the CPT-violating asymmetry in the decay $\Bs \to \mu^\pm D_s^\pm$  
	as a function of celestial direction and 
	sidereal phase. 
	We find no evidence for  CPT-violating effects  and place
	limits on the direction and magnitude of flavor-dependent  CPT- and
	Lorentz-invariance violating coupling coefficients.  
	We find 95\% confidence intervals
	of $\Delta a_{\perp} < 1.2 \times 10^{-12}$~GeV and $(-0.8  <    \Delta
	a_T - 0.396 \Delta a_Z < 3.9) \times 10^{-13}$~GeV. 	
\end{abstract}

\maketitle

Lorentz invariance requires that the description of a particle is
independent of its direction of motion or boost velocity. The Standard
Model Extension (SME)~\cite{CPT2} provides a framework for potential
Lorentz and CPT invariance violation (CPTV), suggesting that such
violations can occur at the Planck scale but still result in potentially
observable effects at currently available collider energies. The process
of neutral meson oscillations is described by a $2 \times 2$ effective
Hamiltonian with mass eigenvalues of the propagating particles having
very small differences between them that drive the oscillation
probability. For the $\Bs$-$\barBs$ system, the fractional difference
between the eigenvalues is of the order of $10^{-12}$. Due to this,
$\Bs$-$\barBs$ oscillations form an interferometric system that is very
sensitive to small couplings between the valence quarks and a possible
Lorentz-invariance violating field, making it an ideal place to search
for new physics~\cite{CPT1}.

The measurement of the like-sign dimuon asymmetry by the D0
Collaboration~\cite{dimuon2013} shows  evidence of anomalously large
CP-violating effects. This is currently one of the few significant
deviations from the standard model of particle physics. One of the
interpretations of this effect could be a CPT-invariant CP violation
(CPV) in neutral $B$-meson mixing. The propagating ``light" ($L$) and
``heavy" ($H$) mass eigenvalues of the $\Bs$-$\barBs$ system can be
written as~\cite{CPT3}:
\begin{align}
|B_{sL} \rangle \propto \, \, & p\sqrt{1-\xi_s}|\Bs \rangle + q\sqrt{1+\xi_s} | \barBs \rangle, \\
|B_{sH} \rangle \propto \, \, & p\sqrt{1+\xi_s}|\Bs \rangle - q\sqrt{1-\xi_s} | \barBs \rangle.
\end{align}
If the complex parameter $\xi_s$ is zero, CPT is conserved and CPV is
due to $|q/p| \ne 1$ so that the oscillation probability $P(\Bs
\rightarrow \barBs)$ is different from $P(\barBs \rightarrow \Bs)$. An
alternative interpretation is that the asymmetry could arise from
T-invariant CPV in $\Bs$-$\barBs$ mixing~\cite{vanKooten} where
$|q/p| = 1$, but $\xi_s$ is non-zero so that the probability of
non-oscillation or oscillation back to the original state $P(\Bs
\rightarrow \Bs)$ is different from $P(\barBs \rightarrow \barBs)$. By
integrating these two probabilities in time the asymmetry $\mathcal{A}_{\mathrm{CPT}}$
between \Bs\ and \barBs\ meson decays can be investigated. 
It can be shown that the CPTV contributions to the
 $2 \times 2$ effective Hamiltonian governing $\Bs$-$\barBs$
oscillations depend on the difference between the diagonal mass and
decay rate terms~\cite{CPT3}:
\begin{align}
\xi_s = \frac{(M_{11} - M_{22}) -  \frac{i}{2}(\Gamma_{11} - \Gamma_{22})}{-\Delta m_s + i\Delta\Gamma_s/2} & \approx \frac{-\beta^{\mu} \Delta a_{\mu}}{\Delta m_s - i \Delta\Gamma_s/2},
\end{align}
where $\Delta a_{\mu}$ is a four vector direction and magnitude in
space-time characterizing Lorentz-invariance violation which in 
the SME is given by $\Delta a_{\mu} = r_{s} a_\mu^{s} - r_{b} a_\mu^{b}$  
where $a_\mu^{q}$ 
are Lorentz-violating coupling constants for the two valence quarks in the \Bs\ meson, 
and where the factors $r_{q}$ allow for quark-binding or other normalization effects.
The four-velocity of the \Bs\ meson is given by  $\beta^{\mu} =
\gamma(1,\vec{\beta})$, 
$\beta^{\mu}\Delta a_{\mu}$ is the 
difference between the diagonal elements of the effective Hamiltonian,
and the  mass and decay rate differences of the mass eigenstates are
$\Delta m_s = m_H - m_L$, and $\Delta \Gamma_s =
\Gamma_L - \Gamma_H$~\cite{convention}. The
small fractional values of $\Delta m_s$ and $\Delta\Gamma_s$  make the \Bs\ system sensitive  to  CPTV
effects. In the underlying theory, spontaneous Lorentz symmetry breaking
generates constant background expectation values for the quark fields
that are Lorentz vectors represented by  
$\Delta a_{\mu}$  or tensors  instead of  scalars~\cite{CPT3}.

Any observed CPT violation should vary in the  frame  of the detector denoted
with indices $(t,x,y,z)$. The period will be one sidereal day ($\simeq
0.99727$ solar days) as the  direction of the proton beam 
follows the Earth's rotation with respect to the distant
stars~\cite{CPT3}.   In the SME the variation would depend on
CPT- and Lorentz-invariance violation coupling coefficients $\Delta
a_{\mu}$ with indices $(T,X,Y,Z)$. 
We choose $(T,X,Y,Z)$ as coordinates 
in the standard Sun-centered frame with the rotation 
axis of the Earth taken
as the $Z$-axis and  $X$($Y$) is at right ascension $0^\circ$ ($90^\circ$)~\cite{Kostelecky:1999mr} (see~\cite{appendix} for a diagram of the coordinate system).
If  CPTV in $\Bs$-$\barBs$
oscillations is allowed, then  $\mathcal{A}_{\mathrm{CPT}}
= (\Delta m_s/\Gamma_s) \mathrm{Im}(\xi_s)$ if $\xi_s$ is small. By
translating from the Sun-centered frame to  the detector frame we
have~\cite{CPT3}
\begin{align}
\mathcal{A}_{\mathrm{CPT}} = &  
{} \frac{-\Delta\Gamma_s \gamma^{_{\mathrm{D0}}}}{\Gamma_s \Delta m_s}
					\left[	\Delta a_T -   C_\alpha S_\chi  \beta_z^{_{\mathrm{D0}}} \Delta a_Z   \right. \nonumber \\
					& +  \left. \sqrt{C^2_\alpha C^2_\chi + S^2_\alpha} \sin(\Omega\hat{t} + \delta +\kappa) \beta_z^{_{\mathrm{D0}}} \Delta a_\perp \right],
\label{eq:3}
\end{align}
where $C_x = \cos(x)$, $S_x = \sin(x)$, $\hat{t}$ is elapsed time with
respect to the vernal equinox of the year 2000, $\Omega =
2\pi$~rad/sidereal day, $\beta_z^{_{\mathrm{D0}}} = \beta^{_{\mathrm{D0}}}\cos\theta$ is
the velocity $\vec{\beta}$ of the \Bs\ meson in the detector frame
projected onto the $z$-axis (proton beam direction) of the D0 detector,
$\theta$ is the polar angle between the \Bs\ momentum and the proton
beam direction, $\gamma^{_{\mathrm{D0}}} = 1/\sqrt{1 -
(\beta^{_{\mathrm{D0}}})^2}$, $\chi$ is the colatitude of the D0
detector,  $\alpha$ is the orientation of the $z$-axis of the detector 
 in the earth's coordinate system, where the proton beam 
has a bearing of 219.53$^\circ$, $\Delta a_{\perp} = \sqrt{\Delta a_X^2+\Delta
a_Y^2}$  is the transverse and $\Delta a_Z$ the longitudinal components of
$\Delta {a}_\mu$, $\delta = \tan^{-1}(\Delta a_Y/\Delta a_X)$, $\kappa =
\tan^{-1}(-S_\alpha/C_\alpha C_\chi)$ and $\Delta a_T$ is the time
component of the $\Delta a_{\mu}$ four-vector. A variation with sidereal time
could arise  from the rotation of $\beta_z^{_{\mathrm{D0}}}$ with
respect to $\Delta \vec{a}$. In this Letter we  place limits on $\Delta a_\perp$
and $\Delta a_T - C_\alpha S_\chi  \beta_z^{_{\mathrm{D0}}} \Delta a_Z$.

Past experiments and analyses have placed constraints on the flavor-dependent $\Delta
a_{\mu}$ in other neutral meson oscillating systems:
$K^0$-$\bar{K}^0$~\cite{K0limits},  $D^0$-$\bar{D}^0$~\cite{D0limits},
and $B^0$-$\bar{B}^0$~\cite{B0limits}, as well as indirect limits for
$B^0_s$-$\bar{B}^0_s$~\cite{vanKooten}.

This article presents a search for CPT and Lorentz violation using the
decay $\Bsdecay X$, where \DsMdecay\ and $\phi \to K^+K^-$ (charge
conjugate states are assumed in this article). CP-violating asymmetries
are usually between ``wrong-sign'' decays $\Bs \to \barBs \to
\mu^+\Dsm$, but we want to study  the asymmetry between the ``right-sign'' decays
$\Bs \to \Bs \to \mu^-\Dsp$ and its charge conjugate. We extract the
CPT-violating parameter using  the asymmetry: 
\begin{equation}
 \label{raw}
	A = \frac{ N_{+} -  N_{-}}{ N_{+} + N_{-}},
\end{equation}
where  $N_{+}$ [$N_{-}$] is the number of reconstructed $\Bs \to \mu^\pm
D_s^{\mp} X$ events where  $\mathrm{sgn}(\cos\theta)Q>0$
[$\mathrm{sgn}(\cos\theta)Q<0$] which results from the
$\beta_z^{{_\mathrm{D0}}} = \beta^{_{\mathrm{D0}}} \cos \theta$ terms in
Eq.~\ref{eq:3} and $Q$ is the charge of the muon. 
The direction of the $\mu^{+} D_s^{-}$ system differs from that of
the parent \Bs\ due to the missing neutrino.   However, the migration
between $N_+$ and $N_-$ terms near $\theta = \pi/2$ causes a negligible
correction to the measured asymmetry.
The initial state at
production is not flavor tagged in our study, but after experimental
selection requirements, the \Bs\ system is fully mixed, so that the
probability of observing a \Bs\ or \barBs\ is essentially equal
regardless of the flavor at production. We assume no CP violation in
mixing~\cite{hfag}, 
so only about half of the observed \Bs\ have the same flavor as they
had at birth.  We assume no CP violation, so those observed \Bs\
mesons which have changed their flavor do not contribute to CPTV,
leading to a ~50\% dilution in the measured asymmetry.
In
the presence of CPT violation, the asymmetry is expected to have a
period of one sidereal day, so a search is made for variations of the
form 
\begin{equation}
A(\hat{t}) =  A_0 - A_1\sin(\Omega \hat{t} + \phi), 
\label{eq:4}
\end{equation}
where $A_0$, $A_1$ and $\phi = \delta + \kappa$ are constants and are
extracted by measuring the asymmetry $A$ in Eq.~\ref{raw} in bins of the
sidereal phase $\Omega\hat{t}$, and fitting to the value in each bin
with Eq.~\ref{eq:4}. Measurements of $A_0$ and $A_1$ are then
interpreted as limits on $\Delta a_{\mu}$ from $\Bs$-$\barBs$
oscillations. A non-zero value of $\Delta a_z$ and $\Delta a_T$ would
lead to a  CPTV asymmetry that does not vary with sidereal time. 

The data selection and the signal extraction are identical to those used
in Ref.~\cite{d0assl}. The main details of the data selection using the D0 detector~\cite{d0det} are
described here.

The data are collected with a suite of single and dimuon triggers. The
selection and  reconstruction of  $\mu^{+} D_s^{-}X$ decays require
tracks with at least two hits in both the central fiber tracker and the
silicon microstrip tracker. The muon track segment outside the
calorimeter has to be matched to a particle found in the central
tracking system which has momentum $p >  3$~GeV and transverse momentum
$2 < p_T < 25$~GeV.
The  $D_s^- \rightarrow  \phi \pi^-$, $\phi \rightarrow K^+ K^-$ decay
is reconstructed by assuming the two $\phi$ decay particles are kaons, 
requiring  $p_T > 0.7$~GeV, opposite
charges, and  $M(K^+K^-) < 1.07$~GeV. The charge of the third
particle, assumed to be the charged pion, must have charge opposite to
that of the muon and $0.5 < p_T<25$~GeV. The three tracks are combined
to create a common $D_s^-$ decay vertex using the algorithm described in
Ref.~\cite{vertex}.  The reconstructed $\mu^\pm D_s^\mp$ candidate is
required to pass several kinematic selection criteria and satisfy
likelihood ratio criteria that are identical to those described in
Ref.~\cite{d0assl}.

The effective  $K^+K^-\pi^\pm$  mass distribution is fitted using bins
of 6~MeV over a  range of $1.7 <  M(K^+K^-\pi^\pm) < 2.3$~GeV, and the
number of signal and background events is extracted by a $\chi^2$ fit of
an empirical model to the data. The $D_s^\pm$ meson mass distribution is
well modeled  by two Gaussian functions constrained to have the same
mean, but with different widths and  normalizations. There is negligible
peaking background under the \Ds\ peak. A second peak in the
$M(K^+K^-\pi^\pm)$ distribution corresponding to the Cabibbo-suppressed
$D^\pm \rightarrow \phi \pi^{\pm}$ decay  is  also  modeled by two
Gaussian functions with widths set to those of the \Ds\ meson model
scaled by the ratio of the fitted $D^\pm$ and \Ds\ masses. The
combinatoric background  is modeled by a $5^{\rm th}$-order  polynomial
function. Partially reconstructed decays such as $\Ds \to \phi \pi^\pm
\pi^0$ where the $\pi^0$ is not reconstructed are modeled with a
threshold function that extends to the \Ds\ mass after the $\pi^0$ mass
has been subtracted, given by $T(m) = \tan^{-1}\left[p_1  (m c^2 - p_2 )
\right] + p_3$, where $p_i$ are fit parameters.

The raw asymmetry  (Eq.~\ref{raw}) is extracted by fitting the
$M(K^+K^-\pi^\pm)$ distribution of the $\mu^\pm D_s^\mp$ candidates
using a $\chi^2$ minimization. The fit is performed simultaneously,
using the same models, on the sum and the difference of the
$M(K^+K^-\pi^\pm)$ distribution of  $N_{+}$ candidates and $N_{-}$
candidates. The functions used to model the two distributions are 
\begin{align}
W_{\rm{sum}} = & W_{D_s}  + W_D + W_{\rm{cb}} + W_{\rm{pt}}, \label{Wsum}\\
W_{\rm{diff}} = &  A W_{D_s} + A_{D}W_{D} + A_{\rm{cb}}W_{\rm{cb}} +A_{\rm{pt}}W_{\rm{pt}},
\end{align}
where $W_{D_s}, W_D$, $W_{\rm{cb}}$, and $W_{\rm{pt}}$ describe the
distribution of the \Ds\ and $D^\pm$ mass peaks, the combinatorial
background, and the partially reconstructed events, respectively, and the $A$ factors are the
corresponding asymmetries which are extracted from the fit.
The number of signal
events in the sample is $N(D_s^\pm) = 205,\!865 \pm 626$.

Following previous conventions~\cite{CPT4}  we
shift the origin of the time coordinate to correspond to the vernal
equinox of the year 2000. The value of $A_1$ is extracted by dividing
the data into $n$ data sets, each containing a fraction $f_i$ of the
data based on the sidereal phase $\Omega\hat{t} + \phi$. In the fit, the
parameters that describe the mass distributions $W_{\mathrm{sum}}$ and $W_{\mathrm{diff}}$ 
are the same for all sidereal bins, except for $A$
and $A_D$ which may vary with sidereal phase.

\begin{figure}[htb]
        \includegraphics[width=\columnwidth]{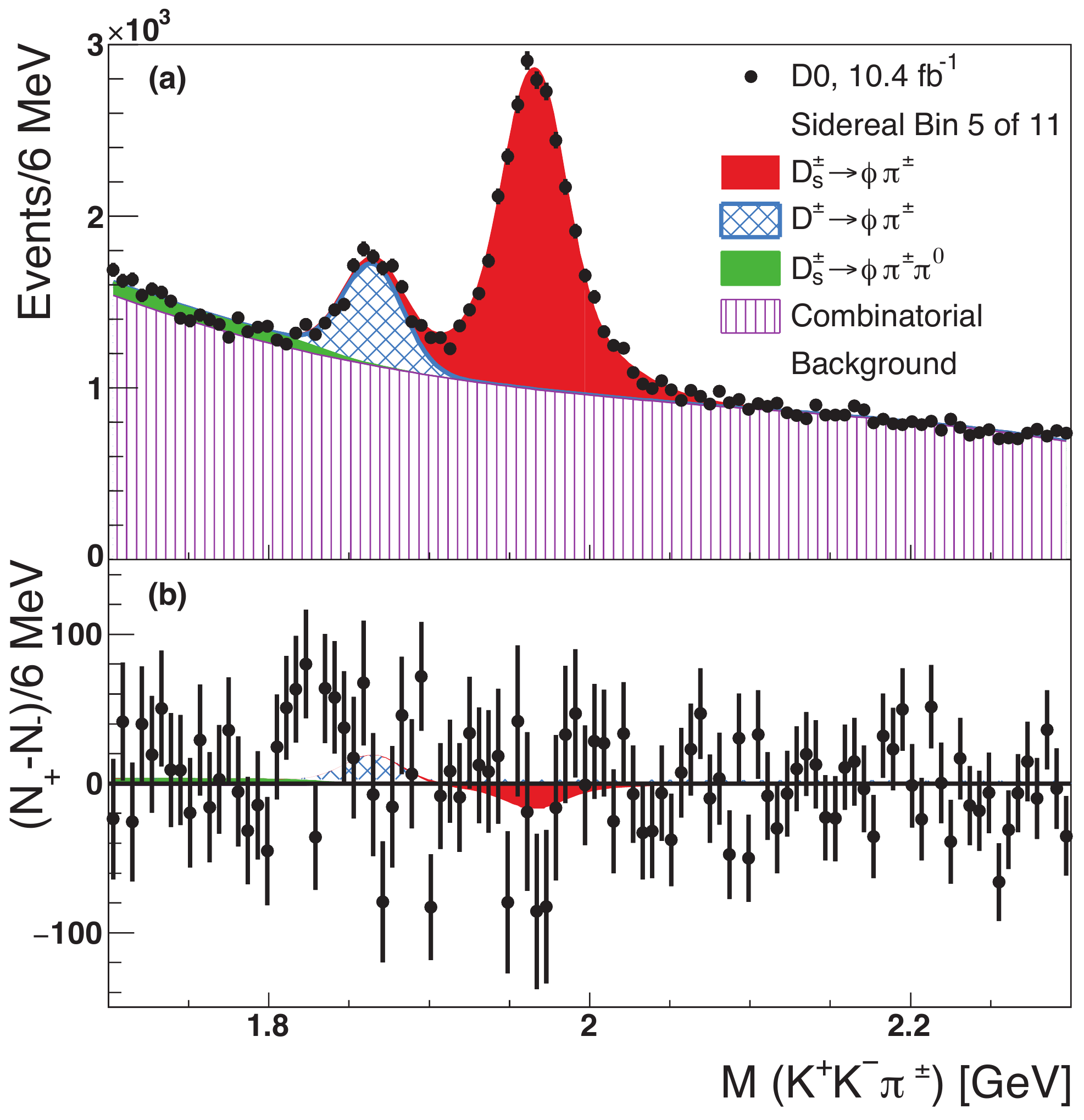}
        \caption[]{(a) The $K^+K^-\pi^\mp$ invariant mass distribution for one of the 11
sidereal bins of the data (bin 5) of the $\mu^{\pm}\phi\pi^\mp$  sample. 
The lower mass peak is due to  the decay
$D^{\mp} \rightarrow \phi \pi^\mp$ and the second peak is due to the
$D_s^{\mp}$ meson decay. (b) The fit  to the $(N_+ - N_-)$ distribution
for one of the 11 sidereal bins of the data (bin 5). 
        }
\label{fig:FourSiderealBins}
\end{figure}

The number of sidereal bins used to extract the asymmetry is determined
by finding the smallest uncertainty on $A_1$. By using MC input of
asymmetries from $0\%$ to $2\%$ we find that the optimum
number of bins  is eleven. One of the eleven 
distributions produced in the fit to the data is shown in
Fig.~\ref{fig:FourSiderealBins}. 

Systematic uncertainties of the fitting method on the extracted values
of $A$ in sidereal bin $i$, $A(i)$,  are evaluated by varying the fitting procedure and are
assigned to be half of the maximal variation in the asymmetry. The mass
range of the fit is shifted from $1.700 <  M(K^+K^-\pi^\pm) < 2.300$~GeV
to $1.724 <  M(K^+K^-\pi^\pm) < 2.270$~GeV in steps of 6~MeV   resulting
in an absolute uncertainty on the measured asymmetries of $0.035\%$. 
The width of the mass bins is changed between 1 and 12~MeV resulting in
an absolute uncertainty of $0.071\%$. The functions modelling the signal
are modified to fit the $D^\pm$ and $D_s^\pm$ mass peaks by single
Gaussian functions, the background is fitted by varying between a
fourth- and seventh-order polynomial function, and the parameter $p_1$
in the threshold function is allowed to vary. As a test, the fraction of
data in each sidereal bin, $f_i$ is fixed to exactly $1/11$. These variations of the signal modelling  yield
an absolute uncertainty on the asymmetry of $0.085\%$. The  uncertainty 
for each of these sources is added in quadrature, to give the total
systematic uncertainty of the fitting procedure of $0.12\%$.  This
uncertainty on the measured values of  $A(i)$ is found to be independent
of sidereal bin, and is added in quadrature to the statistical
uncertainty to extract the CPT-violating parameters by fitting to
Eq.~\ref{eq:4} (see Table~\ref{ConversionNumbers}). The measured values
of the asymmetries, $A(i)$, are plotted in Fig.~\ref{Fig:AsymFit1} and are tabulated in \cite{appendix}.

\begin{figure}[htb]
\includegraphics[width=\columnwidth]{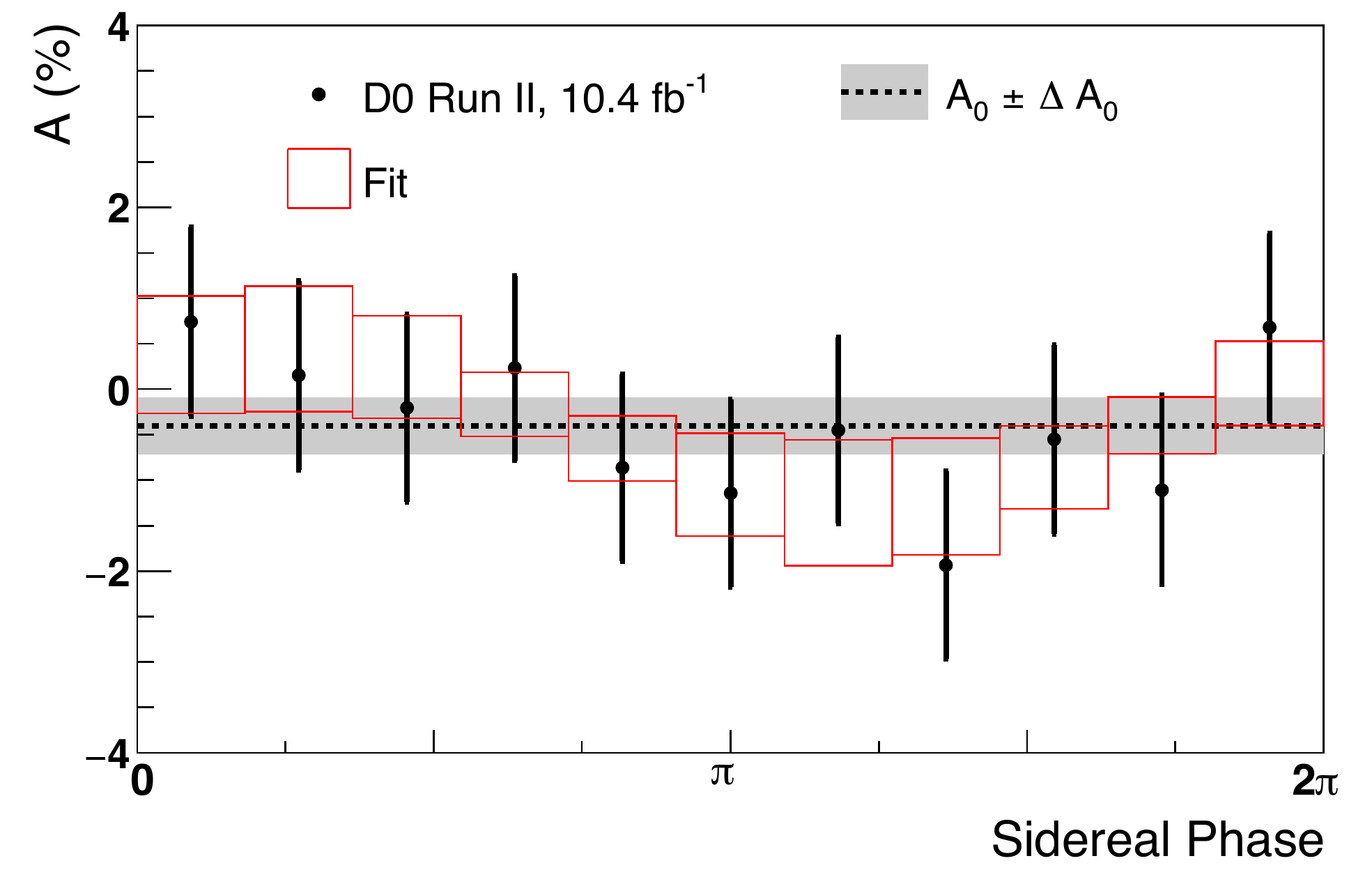}
\caption{
The measured asymmetries, $A(i)$ versus sidereal phase. The uncertainty
on each value of $A(i)$ is the sum in quadrature of the statistical and
systematic uncertainties. The red boxes show the fit and its
uncertainties to the data points (Eq.~\ref{eq:4}). The dashed line shows
the extracted value of $A_0$ and the grey box shows $\Delta A_0$. 
}
\label{Fig:AsymFit1} 
\end{figure}
The limits on $\Delta a_{\mu}$ are extracted using:
\begin{align} 
A_1 & \sin(\Omega \hat{t} + \phi) = 
\frac{{F_{\Bs}^{\text{non-osc}}} \Delta\Gamma_s \langle \gamma^{_{\mathrm{D0}}} \beta_z^{_{\mathrm{D0}}} \rangle}{\Gamma_s\Delta m_s} \ \nonumber \\
& \hspace*{0.8cm} \times \sqrt{C^2_\alpha C^2_\chi + S^2_\alpha} \sin(\Omega\hat{t} + \delta +\kappa) \Delta a_{\perp}, \\
 A_0  = & {-}   \frac{F_{\Bs}^{\text{non-osc}}\Delta\Gamma_s\langle \gamma^{_{\mathrm{D0}}} \rangle }{\Gamma_s\Delta m_s } 
\left[ \Delta a_T -   C_\alpha S_\chi  \langle \beta_z^{_{\mathrm{D0}}}\rangle  \Delta a_Z \right],
\end{align} 
where  angle brackets denote average values. The $F_{\Bs}^{\text{non-osc}}$ factor is the fraction of \Dsdecay\
decays for which an observed \Bs\ has the same flavor as at birth~\cite{d0assl}.
Combining  the fraction of \Bs\ decays in
the sample and the 50\% dilution factor described earlier gives
$F_{\Bs}^{\text{non-osc}} = 0.465$. 
Limits are extracted from the probability distribution which is given by
$\exp({-\chi^2}/{2})$ where $\chi^2$ is
the chi-square as a function of $A_1$, $A_0$ and $\delta$ using Eq.~\ref{eq:4}.
 Since we are setting limits, the probability
distribution will be characterized by two quantities, the most probable
value of $A_1$ and the 95\% upper limit (UL) which is extracted by
integrating the normalized probability distribution at the value of $\delta$
that gives the most conservative limit.

To extract limits, we  measure the average values of $\langle
\gamma^{_{\mathrm{D0}}} \rangle =  {\langle E_{\Bs}  \rangle}/{  m_{\Bs}
}$, $\langle  \beta_z^{_{\mathrm{D0}}}  \rangle= {\langle p_z
\rangle}/{\langle E_{\Bs}  \rangle}$ and $\langle
\gamma^{_{\mathrm{D0}}} \beta_z^{_{\mathrm{D0}}}  \rangle= {\langle p_z
\rangle}/{m_{\Bs}}$ where $\langle p_z \rangle$ is the average momentum
in the $z$-direction and $\langle E_{\Bs}  \rangle$ is the average
energy of the \Bs\ meson. The average momentum of the $\mu\Ds$
candidates is measured using sideband subtraction. The signal region is
$1.92 < M(K^+K^-\pi^-) < 2.00$~GeV and the sideband regions are $1.75 <
M(K^+K^-\pi^-) < 1.79$~GeV and $2.13 < M(K^+K^-\pi^-) < 2.17$~GeV, and
the average is $\langle p \rangle = 21.41 \pm 0.03$~GeV. This momentum
needs to be corrected for the missing neutrino in the decay using a
$k$-factor correction. These $k$-factors are taken from
Ref.~\cite{d0note6148} and applied to give a  momentum of $\langle p
\rangle = 25.3$~GeV. The systematic uncertainty on $\langle p \rangle$
of 1.6~GeV is obtained from the difference between the momentum
extracted using sideband subtraction and using a weighted average of the
number of signal events in  momentum bins 
which is then added in quadrature to the uncertainty due to the $k$-factors.
The effect of possible
reconstruction variations in the $x$ and $y$ directions are found to be
less than 1\%. If we vary the number of sidereal bins the most probable
value of $A_1$ varies by 8\%. These variations are added in
quadrature as the relative systematic
uncertainty on the value of $A_1$.

The final results are obtained by scaling the probability distributions
obtained for $A_0$, $A_1$ with the multiplicative factors given in
Table~\ref{ConversionNumbers}. The systematic uncertainties on the
multiplicative factors, the number of sidereal bins, and reconstruction
effects are included by convoluting the probability distribution with a
Gaussian function with the width given by the sum in quadrature of the
systematic uncertainties. We obtain a 95\% upper limit (UL) of $\Delta
a_{\perp} < 1.2 \times 10^{-12}$~GeV. The most probable values of
$\delta$ and $\Delta a_{\perp}$ are $\delta = 4.901$ and $\Delta
a_{\perp} = 5.7\times 10^{-13}$~GeV.

\begin{table}[htb]
\caption{Parameters and  uncertainties in the extraction of the 
CPT-violating parameters. The uncertainties on $A_0$, $A_1$ and $\phi$ are fit uncertainties and are dominated by the statistical uncertainty of the raw asymmetries. All other uncertainties are systematic.
}
\begin{center}
\begin{ruledtabular}
\newcolumntype{A}{D{A}{\pm}{-1}}
\begin{tabular}{ccc}
Parameter & \multicolumn{1}{c}{Value} & Ref. \\ 
\hline
$A_0$ &  $(-0.40 \pm 0.31)\%$ & Eq.~\ref{eq:4} \\
$A_1$ &  $(0.87 \pm 0.45)\%$ &  Eq.~\ref{eq:4}\\
$\phi$ & $-2.28 \pm 0.51$ & Eq.~\ref{eq:4}\\
 \hline 
$m_{\Bs}$ 		& $(5.36677 \pm  0.00024) \, \mathrm{GeV}$ & \cite{pdg} \\
$\Delta m_s$ & $(17.761 \pm 0.022) \times 10^{12} \, \hbar \mathrm{s}^{-1}$ & \cite{pdg}\\
$\Delta \Gamma_s / \Gamma_s$    & $(0.138 \pm 0.012)$ & \cite{pdg}\\
$F_{\Bs}^{\text{non-osc}} = F_{\Bs}^{\text{osc}}$ & $(0.465 \pm 0.017)$ & \cite{d0assl}\\
$\langle p_z \rangle$ 		& $(17.8 \pm 1.6)\thinspace\mathrm{GeV}$ & \\
$\langle p \rangle$ 		& $(25.3 \pm 2.3)\thinspace\mathrm{GeV}$ &  \\
Proton beam dir$^{\mathrm{n}}$ $\alpha$		& 	219.53$^\circ$	& \\
Colatitude $\chi$ & $48.17^\circ$ & \\ 
\end{tabular}	
\end{ruledtabular}
\end{center}
\label{ConversionNumbers}
\end{table}%

The limit on $\Delta a_T -   C_\alpha S_\chi \beta_z^{_{\mathrm{D0}}}
\Delta a_Z$  is obtained from a  fit to the asymmetries using
Eq.~\ref{eq:4}. This results in a value of $A_0 = (-0.40 \pm 0.31)\%$. 
In this case the systematic uncertainties on the measured values of
$A(i)$ are assumed to be 100\% correlated between sidereal bins to
obtain the most conservative limits and are added to the statistical
uncertainty obtained from the fit. Using Eq. 10, we obtain  $\Delta a_T
- C_\alpha S_\chi  \beta_z^{_{\mathrm{D0}}} \Delta a_Z = \Delta a_T -
0.396 \Delta a_Z = (1.5 \pm 1.2)\times 10^{-13}$~GeV resulting in  a two
sided 95\% confidence interval $(-0.8  <    \Delta a_T - 0.396 \Delta
a_Z < 3.9) \times 10^{-13}$~GeV.

We did a cross check using the periodogram methodology~\cite{period} which sees no anomalous behavior for the frequency 1/sidereal day \cite{appendix}.

For CPTV to explain the  difference between   the like-sign dimuon
asymmetry~\cite{dimuon2013} and the SM requires that $ (\Delta a_T -
0.396 \Delta a_Z)$ to be of the order of
$10^{-12}$~GeV~\cite{vanKooten}. These limits imply that CPT violation
is unlikely to  contribute a significant fraction of the observed dimuon
charge asymmetry, and that other explanations need to be sought.

In conclusion, we have carried out the first search for CPT-violating
effects exclusively in the $\Bs$-$\barBs$ oscillation system via
semileptonic decays of the \Bs\ mesons. We find no significant evidence
for  CPT-violating effects and place limits on the size of the Lorentz
violating effects, $\Delta a_{\mu}$. These limits constrain a linear combination of the Lorentz-violating coupling constants $a^q_{\mu}$ for the $b$ and $s$ valence quarks in the \Bs\ meson that are different from the linear combinations of valence quarks in the $B^0$~\cite{B0limits} or $K^0$~\cite{K0limits} mesons, and therefore further constrain the possible separate values of the three coefficients $a^b_{\mu}$, $a^s_{\mu}$, and $a^d_{\mu}$.
 We find 95\% confidence intervals
for the flavor-dependent coefficients 
 $\Delta a_{\perp} < 1.2 \times 10^{-12}$~GeV and $(-0.8  < \Delta
a_T - 0.396 \Delta a_Z < 3.9) \times 10^{-13}$~GeV.

\begin{acknowledgements}
We thank A. Kosteleck\'y for valuable conversations during the course of this work.
We also thank the staffs at Fermilab and collaborating institutions,
and acknowledge support from the
Department of Energy and National Science Foundation (United States of America);
Alternative Energies and Atomic Energy Commission and
National Center for Scientific Research/National Institute of Nuclear and Particle Physics  (France);
Ministry of Education and Science of the Russian Federation, 
National Research Center ``Kurchatov Institute" of the Russian Federation, and 
Russian Foundation for Basic Research  (Russia);
National Council for the Development of Science and Technology and
Carlos Chagas Filho Foundation for the Support of Research in the State of Rio de Janeiro (Brazil);
Department of Atomic Energy and Department of Science and Technology (India);
Administrative Department of Science, Technology and Innovation (Colombia);
National Council of Science and Technology (Mexico);
National Research Foundation of Korea (Korea);
Foundation for Fundamental Research on Matter (The Netherlands);
Science and Technology Facilities Council and The Royal Society (United Kingdom);
Ministry of Education, Youth and Sports (Czech Republic);
Bundesministerium f\"{u}r Bildung und Forschung (Federal Ministry of Education and Research) and 
Deutsche Forschungsgemeinschaft (German Research Foundation) (Germany);
Science Foundation Ireland (Ireland);
Swedish Research Council (Sweden);
China Academy of Sciences and National Natural Science Foundation of China (China);
and
Ministry of Education and Science of Ukraine (Ukraine).
We also acknowledge support from the Indiana University Center for Spacetime Symmetries (IUCSS).
\end{acknowledgements}

\newpage

\section*{Auxiliary material}
\begin{center}
To appear as an Electronic Physics Auxiliary Publication (EPAPS)
\end{center}
\twocolumngrid

\section{Coordinate System}

\begin{figure}[hb!]
\subfigure[]{{\includegraphics[width=0.35\linewidth]{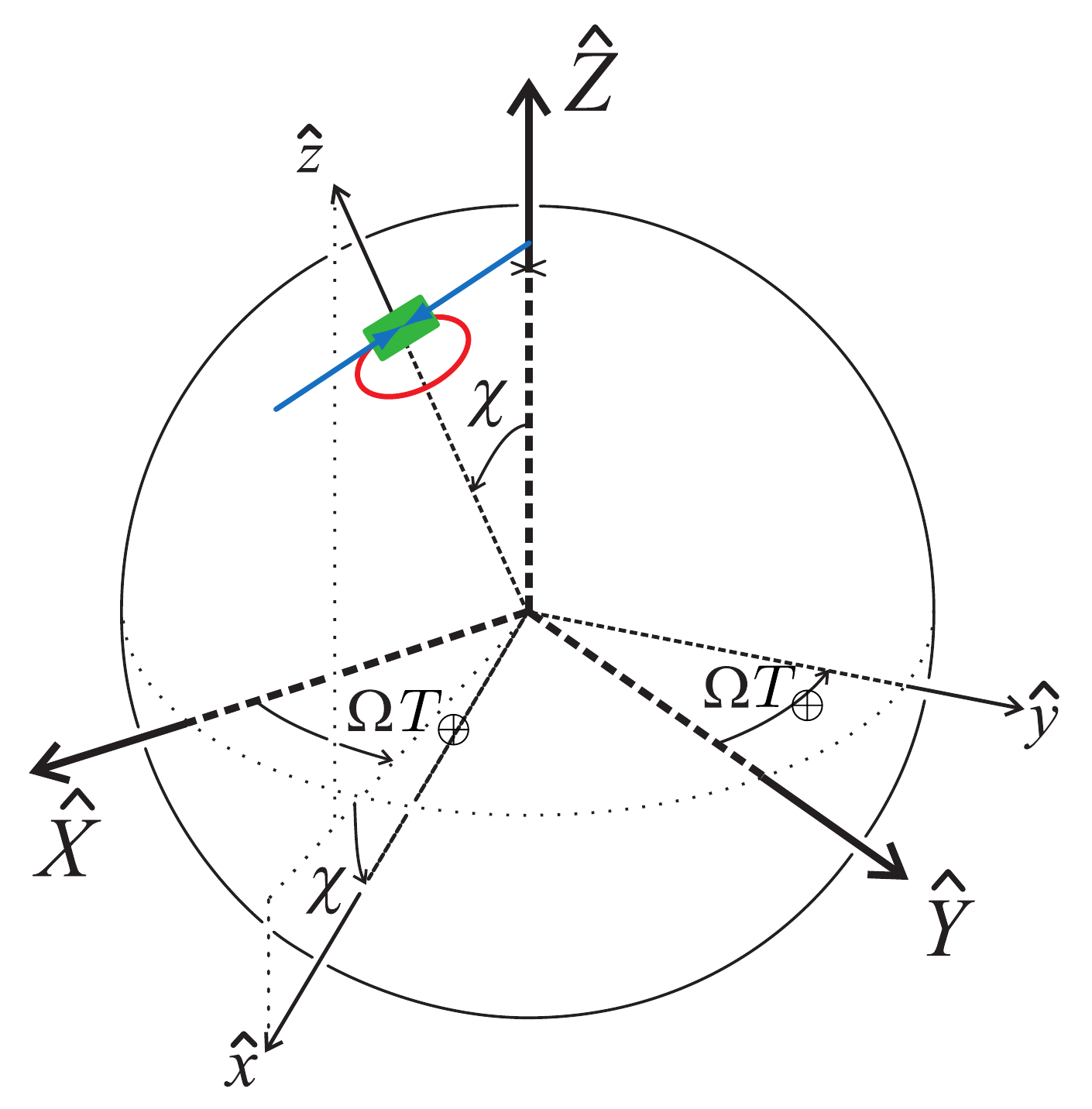}}}
\subfigure[]{\includegraphics[width =0.575\columnwidth]{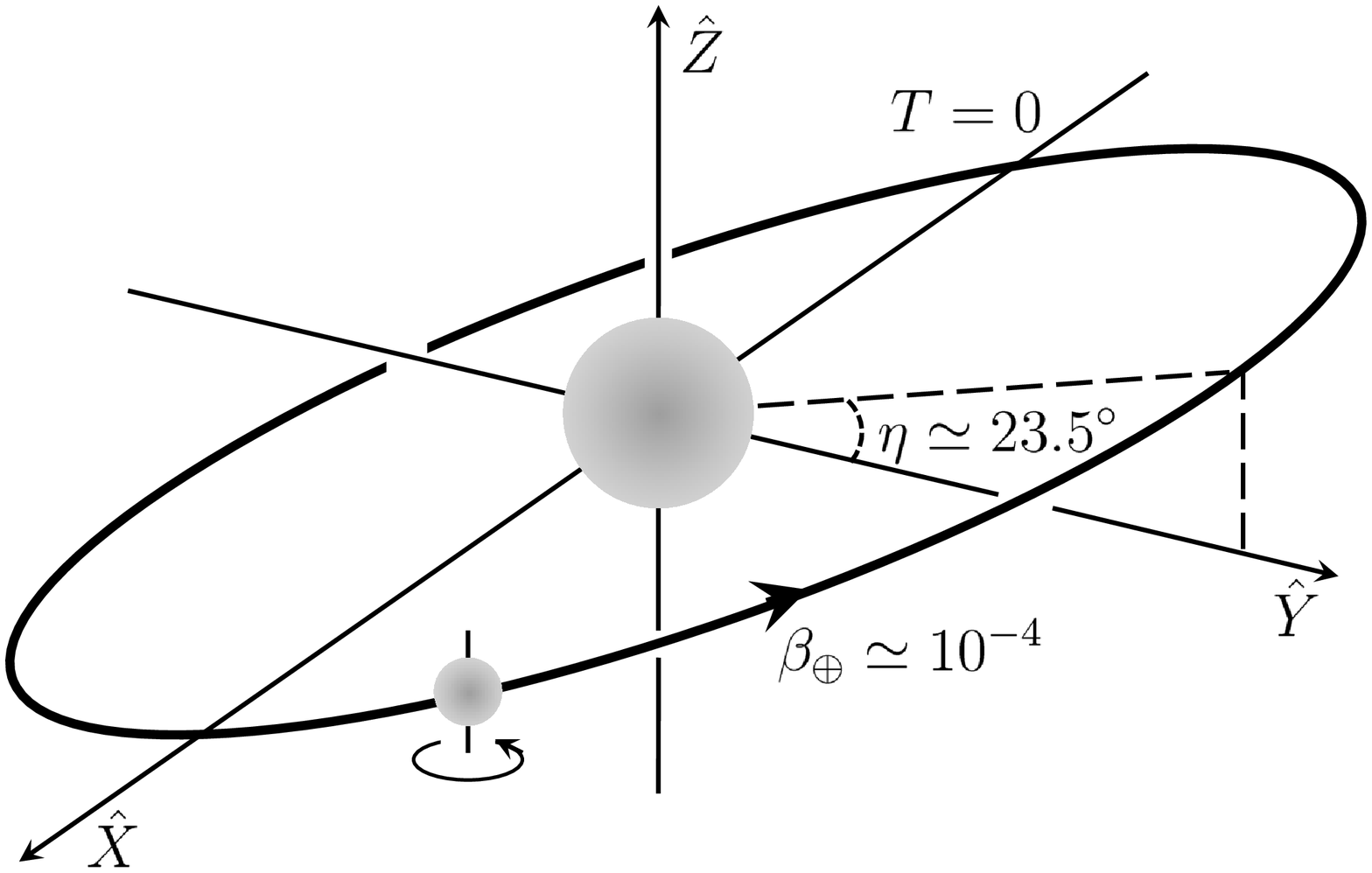}}
\caption{\label{Fig:eorbit} 
 Illustrations of the coordinate systems used in this analysis. (a)  The small rectangle represents the position of the D0 detector on the earth. (b) Orbit of Earth in Sun-based frame (based on Fig.~1 from Ref.~[7]). 
}
\end{figure}

We choose $(T,X,Y,Z)$ as coordinates 
in the standard Sun-centered frame with $T$ being the time coordinate, the rotation 
axis of the Earth taken
as the  choice for the $Z$-axis and  $X$($Y$) is at right ascension $0^\circ$ ($90^\circ$) (c.f. Ref.~[7] of the paper).
This coordinate system is illustrated in Fig.~\ref{Fig:eorbit}.

\section{Measured Asymmetries}	

The measured asymmetries, $A(i)$, used to extract the limits are given in Table~\ref{tab:valiuesOfA}.

\begin{table}[htbp]
\caption{\label{tab:valiuesOfA}
 The measured asymmetries, $A(i)$ versus sidereal phase. The uncertainty
on each value of $A(i)$ is the sum in quadrature of the statistical and
systematic uncertainties.
 }
\begin{ruledtabular}
\newcolumntype{A}{D{A}{\pm}{-1}}
\newcolumntype{B}{D{A}{\to}{-1}}
\scriptsize
\begin{tabular}{cBA}
Asymmetry &  \multicolumn{1}{c}{Sidereal Phase} & \multicolumn{1}{c}{Value (\%)}\\
 \hline  
$A(1)$ &          0 A\ (2\pi)/11 & +0.74\ A\  1.03 \\
$A(2)$ &  (2\pi)/11 A\ 2(2\pi)/11 & +0.15\ A\  1.03 \\
$A(3)$ & 2(2\pi)/11 A\ 3(2\pi)/11 & -0.20\ A\  1.02 \\
$A(4)$ & 3(2\pi)/11 A\ 4(2\pi)/11 & +0.23\ A\  1.01 \\
$A(5)$ & 4(2\pi)/11 A\ 5(2\pi)/11 & -0.86\ A\  1.02 \\
$A(6)$ & 5(2\pi)/11 A\ 6(2\pi)/11 & -1.14\ A\  1.02 \\
$A(7)$ & 6(2\pi)/11 A\ 7(2\pi)/11 & -0.45\ A\  1.02 \\
$A(8)$ & 7(2\pi)/11 A\ 8(2\pi)/11 & -1.93\ A\  1.03 \\
$A(9)$ & 8(2\pi)/11 A\ 9(2\pi)/11 & -0.55\ A\  1.03 \\
$A({10})$ & 9(2\pi)/11 A\ 10(2\pi)/11 & -1.11\ A\  1.03\\
$A({11})$ & 10(2\pi)/11 A\ (2\pi) & +0.68\ A\  1.03 \\
\end{tabular}
\end{ruledtabular}
\end{table}

\section{Periodogram Analysis}
As a cross check to fitting the data for a periodic signal, we  also use
the periodogram~[18] method to measure the spectral power of a
signal over a large range of frequencies. The spectral power at a test
frequency $\nu$ is
\begin{equation}
P(\nu) \equiv \frac{\left| 
\sum\limits^{N}_{j=1}  w_j \exp\left(-2\pi i \nu \hat{t}_j\right)\right|^2}
{N\sigma^2_w},
\end{equation}
where the data has $N$ measurements each of weight $w_j$ where the
weight is the probability that the event is a signal event with a
variance $\sigma_w$. The weight for each event depends on $Q_{j} \cos\theta_j$, and 
$M(K^+K^-\pi^\pm)$ for the event and is based on
the fit to Eq.~7: $w_j ={Q_{j} \cos\theta_j
W_{D_s}[M(K^+K^-\pi^\pm)]}/{W_{\rm{sum}}[M(K^+K^-\pi^\pm)]}$. In the
absence of an oscillatory signal, the probability that $P(\nu)$ at
frequency $\nu$  would exceed an observed
value $S$ is $ P(\nu) > S = \exp(-S)$.

\begin{figure}[hb!]
\includegraphics[width =0.75\columnwidth]{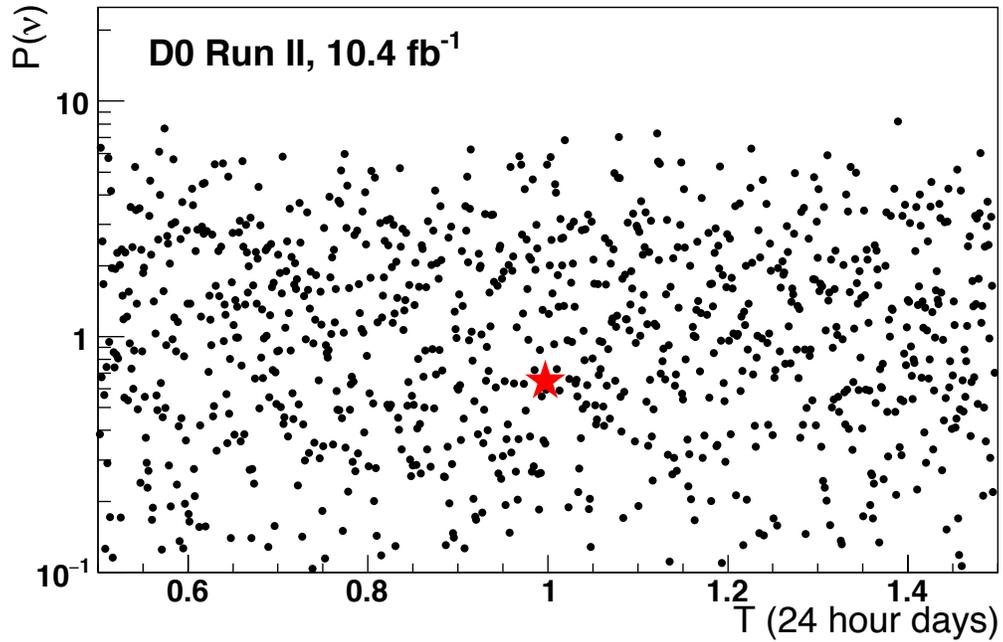}
\caption{\label{Fig:Periodogram_Diff_Bs_2014_06_06} 
The periodogram for the  \Bs\ data sample over the range of 0.5
days to 1.5 days in steps of (1 day/1000). The red
star indicates the spectral power calculated at one sidereal day. 
}
\end{figure}

The spectral power of this data sample is $P(\mathrm{one~sidereal~day})
= 0.65$.  The probability of obtaining a value of $P$ greater than this
is 52\% which is  consistent with no signal. The spectral power values
for periods from 0.5 to 1.5 solar days in steps of 1 solar day/1000 are
shown in Fig.~\ref{Fig:Periodogram_Diff_Bs_2014_06_06}. Sixty percent of
these measurements are greater than the spectral power at one sidereal
day. The 95\% UL is obtained by injecting simulated signals into the
data and determining the probability distribution of the spectral power
as a function of the injected signal $A_1$. The resulting 95\% UL on
$A_1$ is 1.03\%.  This converts to a  95\% UL of $\Delta a_\perp < 6.9
\times 10^{-13}$~GeV which is  comparable to that obtained from the
analysis of the amplitudes.
\onecolumngrid

\begin{thebibliography}{99}
\bibitem{convention} 
This is the convention of the definition of $\Delta \Gamma_s$ so that it is positive in the
standard model.
\end{thebibliography}
\end{document}